\documentclass[journal]{IEEEtran}

\IEEEoverridecommandlockouts

\usepackage{amsmath,amssymb,amsfonts,empheq,bm}
\usepackage{algorithmic}
\usepackage{graphicx}
\usepackage{textcomp}
\usepackage{xcolor}
\usepackage{float}

\usepackage{dotlessi}
\usepackage{layouts}
\usepackage{mathtools}
\usepackage[font=small]{caption}
\usepackage{multirow}
\usepackage{amsmath}
\usepackage{amsfonts}
\usepackage{bm}
\usepackage{etoolbox}
\usepackage{cite}
\usepackage{comment}

\newcommand{\hcoi}{c}
\newcommand{\wcoi}{\omega_{\rm CoI}}
\newcommand{\Wiener}[1]{\bm{w}_{#1}}
\newcommand{\Bbus}{\bm{B}_{\rm bus}}
\newcommand{\Gbus}{\bm{G}_{\rm bus}}
\newcommand{\Ybus}{\bar{\bm{Y}}_{\rm bus}}

\begin{document}

\title{Impact of Power Fluctuations on Frequency Quality} 

\author{%
  \'Angel Vaca, \IEEEmembership{Student Member, IEEE} %
  and Federico Milano, \IEEEmembership{Fellow, IEEE}%
  \thanks{{\'A}.~Vaca and F.~Milano are with the School of Electrical and Electronic Engineering, University College Dublin, Belfield Campus, D04V1W8, Ireland. e-mails: angel.vaca1@ucdconnect.ie, federico.milano@ucd.ie}%
  \thanks{This work is supported by the Science Foundation Ireland (SFI) by funding {\'A}.~Vaca and F.~Milano under NexSys project, Grant No.~21/SPP/3756.}
}%

\maketitle

\begin{abstract}
This paper analyzes how power injections affect frequency quality in power systems.  We first derive a general expression linking active and reactive power injections at buses to the system's frequency.  This formulation explicitly considers both real and imaginary frequency components, providing a complete description of frequency behavior in power systems during transients.  Next, we extend our analysis to incorporate stochastic variations of power injections. Using the frequency divider concept and power-based frequency estimation, we develop analytical relationships linking stochastic load fluctuations to frequency deviations.
We discuss under which conditions the Central Limit Theorem cannot be applied to capture the frequency distribution, thereby clarifying how its hypotheses are not satisfied in power system applications.  Then, we establish clear criteria for the appropriate use of statistical methods in frequency analysis.  Finally, we validate our theoretical results through simulations on modified IEEE 14-bus and all-island Irish transmission test systems, highlighting the accuracy, practical utility, and limitations of our proposed formulation.
\end{abstract}

\begin{IEEEkeywords}
  Stochastic Differential Equations, power fluctuations, frequency quality, noise propagation, Central Limit Theorem, frequency divider, complex frequency.
\end{IEEEkeywords}

\section{Introduction}

\subsection{Motivation}

Conventional frequency analysis in power systems primarily relies on deterministic approaches based on synchronous machine rotor speeds and global frequency indicators, such as the frequency of the center of inertia \cite{TaskForce}.  Recent methods like the Frequency Divider Formula (FDF) provide practical approximations by estimating local bus frequencies directly from machine rotor speeds, proving effective for transient stability analysis and control applications \cite{milano_frequency_2017, milano_method_2021}.  Further advancements include the Complex Frequency (CF) concept, which explicitly considers voltage magnitude and phase-angle dynamics in a unified analytical framework \cite{ComplexFreq}.  

The complex frequency formulation clarifies the dynamic interplay between active and reactive power, becoming particularly relevant in low-inertia and converter-dominated power grids, where accurate transient representation is essential.  However, these deterministic methods do not adequately capture the stochastic nature of modern power systems, as mentioned in \cite{vorobev_deadbands_2019, doheny_investigation_2017}.  In this paper, we propose a formulation that integrates the complex frequency concept with the study of stochastic fluctuations of power injections at network buses.

\subsection{Literature Review}

With the increasing integration of renewable energy sources, distributed generation, and dynamic loads, the analysis of power system frequency has become inherently more complex and uncertain \cite{ kundur, Rezvani, Acha, dorfler2023control, SHAZON20226191}. These new generation resources and variable loads introduce significant stochastic fluctuations, challenging traditional deterministic modeling and analysis methods \cite{schafer_non-gaussian_2018, kato_estimation_2011, martinez-barbeito_dynamical_2023, ma_effect_2024, ghosh_estimation_2025, fernandes_data-driven_2024}.  Consequently, frequency deviations no longer follow purely deterministic or predictable patterns, requiring advanced probabilistic methods to ensure accurate analysis; while some techniques, such as reconciliation, are used to diminish the errors introduced in the estimation of stochastic variables \cite{bai_distributed_2019, sobolewski_estimation_2013}.

In response to these challenges, stochastic modeling techniques have gained prominence for their ability to realistically represent uncertainty in power system dynamics.  In particular, stochastic differential equations (SDEs) provide a robust mathematical toolset to model frequency dynamics under random load variations and renewable generation fluctuations \cite{zhang_derivative_2013, oksendal_stochastic_2000}.  Functional models based on SDEs address renewable variability and dynamic load perturbations, enhancing realism and accuracy in frequency stability analysis \cite{6039188}.

Despite their strengths, many stochastic approaches rely on statistical simplifications, often justified through the Central Limit Theorem (CLT), assuming independence, identical distributions, and thus, Gaussian behavior.  These assumptions frequently remain unverified in practical scenarios and may not hold in large-scale power systems, leading to noticeable inaccuracies or unrealistic simplifications \cite{zhang_central_2023, fante_central_2001, dehay_central_2013}.  As a result, there is a critical need for more precise analytical frameworks that rigorously evaluate the validity and limitations of these statistical approximations.

\subsection {Contributions}

Motivated by these gaps, this paper presents a novel analytical framework, first establishing a generalized deterministic relationship between local power injections and system-wide frequency response using the recently developed CF concept \cite{ComplexFreq}.  Subsequently, we extend this deterministic analysis into a comprehensive stochastic framework, explicitly accounting for random variations in power injections.  Our formulation provides clear mathematical relationships describing how stochastic fluctuations propagate through the network, directly affecting frequency quality.  By employing this analytical framework, we investigate the conditions under which commonly invoked statistical assumptions, particularly those underlying the CLT, remain valid or fail.  Our proposed method allows determining when these approximations can be reliably applied.

\subsection {Paper Organization}

The remainder of this paper is organized as follows.  Section \ref{sec:method} introduces the theoretical foundations and detailed mathematical derivation of the proposed analytical framework.  Section \ref{sec:CLTdescription} leverages this theoretical basis to critically analyze the applicability of the CLT in power systems, highlighting key limitations and offering practical guidelines for its correct application. Section \ref{sec:studycases} validates the proposed framework through numerical simulations on power system models, including modified versions of the IEEE 14-bus test system and the all-island Irish transmission system. Section \ref{sec:finalremarks} briefly discusses broader implications and potential extensions of the method, specifically addressing its application to wind generation reconciliation techniques and analyzing asymmetries in frequency deviations.  Finally, Section \ref{sec:conclusion} summarizes the main conclusions and suggests directions for future research.

\subsection{Notation}

The following notation is utilized.
\begin{itemize}[\IEEEiedlabeljustifyl \IEEEsetlabelwidth{Z} \labelsep 0.9cm]
\item[$a$] scalar quantity
\item[$a(t)$] time-varying quantity
\item[$da$] infinitesimal variation of $a$
\item[$\bar{a}$] complex quantity
\item[$\dot{a}$] time derivative of $a$
\item[$\bm a$] vector
\item[$a_i$] $i$-thv ector element
\item[$\bm a^{\top}$] transpose of $\bm a$
\item[$\bm A$] matrix
\item[$A_{hk}$] $(h,k)$-th matrix element
\item[$\bm A^{-1}$] inverse of $\bm A$
\item[$\bm A^{+}$] Moore-Penrose pseudo-inverse of $\bm A$
\end{itemize}

\section{Derivations}
\label{sec:method}

In this section, we derive the following expressions:
\begin{itemize}
\item $\bm{\omega}(\dot{\bm{p}}, \dot{\bm{q}})$: the dependency of bus voltage frequencies on the rate of change of active and reactive power injection at network buses.
\item $\wcoi(\dot{\bm{p}}, \dot{\bm{q}})$: the dependency of the frequency of the center of inertia (CoI) on the rate of change of active and reactive power injections at network buses.
\item $\wcoi(\dot{\bm{p}})$: an approximated expression of the dependency of frequency of the CoI on the rate of change of active power injections at buses.
\item $d\wcoi(d\bm p, d\bm q)$: the dependency of the variations of the frequency of the CoI on the stochastic fluctuations of active and reactive power injections at buses.
\end{itemize}

\subsection{General Derivation from Complex Frequency Components}

Following the formulation developed in \cite{ComplexFreq}, In this section, we derive a novel expression for the frequencies at the network buses, $\bm{\omega}$, in terms of active and reactive power components at the buses, $\bm{p}$ and $\bm{q}$, respectively.  

To obtain this expression, we consider as starting point the definition of complex frequency given in \cite{ComplexFreq}:
\begin{equation}
    \dot{\bar{\bm v}}(t) = (\bm{\rho}(t) + j \bm{\omega}(t)) \circ \bar{\bm{v}}(t) ,
\end{equation}
where $\bar{\bm{v}} = \bm{v} \angle \bm{\theta}$ is the vector of bus voltage Park vectors, $\bm{\rho}$ and $\bm{\omega}$ are the real and imaginary components of the complex frequency of the voltages, and $\circ$ is the element-by-element product.  For the $i$-th bus, one has:
\begin{align}
   \rho_i(t) = \dot{u}_i(t) = \frac{\dot{v}_i(t)}{v_i(t)}, \qquad \omega_i(t) = \dot{\theta}_i(t) ,
\end{align}
with $u_i = \log(v_i)$.
For an ac transmission grid, the apparent powers injected at network buses can be written as a function of $\bm{v}$ and $\bm \theta$:
\begin{align}
    \dot{\bar{\bm{s}}}(\bm{v}, \bm \theta) &= \dot{\bm{p}}(\bm{v}, \bm \theta) + 
    j \dot{\bm{q}}(\bm{v}, \bm \theta), 
\end{align}
where, for simplicity, we have omitted the explicit dependency on time.  Then, the time derivative of active and reactive power injections leads to, respectively:
\begin{equation}
   \label{eq:dot_active_power}
\begin{aligned}
    \dot{\bm{p}} &= \frac{\partial \bm{p}}{\partial \bm{v}} \, \frac{\partial \bm{v}}{\partial \bm{u}} \, \dot{\bm{u}} + \frac{\partial \bm{p}}{\partial \bm{\theta}} \, \dot{\bm{\theta}} \\ &= \bm{A} \bm{\rho} + \bm{B} \bm{\omega},
\end{aligned}
\end{equation}
and
\begin{equation}
\begin{aligned}
  \label{eq:dot_react_power}
  \dot{\bm{q}} &=  \frac{\partial \bm{q}}{\partial \bm{v}} \, \frac{\partial \bm{v}}{\partial \bm{u}} \, \dot{\bm{u}} + \frac{\partial \bm{q}}{\partial \bm{\theta}} \, \dot{\bm{\theta}} \\
  &= \bm{C} \bm{\rho} + \bm{D} \bm{\omega}.
\end{aligned}
\end{equation}

Solving \eqref{eq:dot_react_power} for $\bm{\rho}$ yields:
\begin{align}\label{eq:rho_equation}
    \bm{\rho} &= \bm{C}^{-1} \left[\dot{\bm{q}} - \bm{D} \bm{\omega}\right].
\end{align}

Substituting \eqref{eq:rho_equation} into \eqref{eq:dot_active_power}, we obtain:
\begin{align}
\dot{\bm{p}} &= \bm{A} \bm{C}^{-1} \dot{\bm{q}} - \bm{A} \bm{C}^{-1} \bm{D} \bm{\omega} + \bm{B} \bm{\omega} \\
&= \bm{A} \bm{C}^{-1} \dot{\bm{q}} + \left( \bm{B} - \bm{A} \bm{C}^{-1} \bm{D} \right) \bm{\omega}.
\end{align}

Defining:
\begin{align}
\bm{E} &= \bm{A} \bm{C}^{-1}, \\
\bm{F} &= \bm{B} - \bm{A} \bm{C}^{-1} \bm{D},
\end{align}
we rewrite the equation as:
\begin{align}
\dot{\bm{p}} &= \bm{E} \dot{\bm{q}} + \bm{F} \bm{\omega},
\end{align}
and solving for $\bm{\omega}$ gives:
\begin{align}
\bm{\omega} &= \bm{F}^{-1} \dot{\bm{p}} - \bm{F}^{-1} \bm{E} \dot{\bm{q}}.
\end{align}

Introducing the shorthands:
\begin{align}
\bm{H} &= \bm{F}^{-1}, \\
\bm{K} &= -\bm{F}^{-1} \bm{E},
\end{align}
and recovering the dependency on time, we obtain the final compact expression:
\begin{equation} \label{eq:coi_bus}
    \boxed{\bm{\omega}(t) = \bm{H}(t) \, \dot{\bm{p}}(t) + \bm{K}(t) \,  \dot{\bm{q}}(t)}
\end{equation}

Detailed derivations and expressions of matrices $\bm{A}$, $\bm{B}$, $\bm{C}$, $\bm{D}$, $\bm{H}$, $\bm{F}$, and $\bm{K}$ are given in the Appendix.

Equation \eqref{eq:coi_bus} provides a direct linear relationship between the power injection derivatives and the local frequency components at each bus.  This relationship is general and only requires that the dynamics of transmission lines are fast enough to be negligible during transients.

\subsection{Frequency of the Center of Inertia}

To extend this formulation to the frequency of the CoI, we follow the framework in \cite{milano_rotor_2018}.  The exact formula of the CoI is, as it is well known:
\begin{equation}
\label{eq:coi}
\wcoi(t) = \frac{\sum_{k=1}^m M_k \omega_{{\rm g}, k}(t)}{\sum_{k=1}^m M_k} = \bm{m}_{\rm g}^{\top} \bm{\omega}_{\rm g}(t) \, ,
\end{equation}
where $M_k$ and $\omega_{{\rm g}, k}$ are the starting time and rotor speed of the $k$-th synchronous machine, respectively; $m$ is the total number of machines; and $\bm{m}_{\rm g}$ is the normalized vector of inertia constants.  From \cite{milano_frequency_2017}, the rotor speeds of the machines are linked to the bus frequencies through the \textit{frequency divider} expression:
\begin{equation}
  \bm B_{\rm bg} (\bm \omega_{\rm g}(t) - \bm 1_m) = [\bm B_{\rm bus} + \bm B_{\rm g}] (\bm \omega(t) - \bm 1_n) \, , 
\end{equation}
where $\bm B_{\rm bus}$ is the $n \times n$ susceptance matrix of the grid; $\bm B_{\rm bg}$ is the $m \times n$ matrix obtained using the stator and step-up transformer impedances of the synchronous machines; $\bm 1_n$ and $\bm 1_m$ are unit vectors of order $n$ and $m$, respectively; and $\bm B_{\rm g}$ is a $n\times n$ diagonal matrix that accounts for the internal susceptances of the synchronous machines and step-up transformers at generator buses.  In \cite{milano_rotor_2018}, it is shown that the Moore-Penrose pseudo-inverse of $\bm B_{\rm bg}$ allows obtaining an explicit expression $\bm \omega_{\rm g}$ in terms of bus frequencies $\omega$, namely:
\begin{equation}
  \label{eq:wb2wg}
  \bm \omega_{\rm g}(t) = \bm B_{\rm bg}^{+} [\bm B_{\rm bus} + \bm B_{\rm g}] (\bm \omega(t) - \bm 1_n) + \bm 1_m \, .
\end{equation}
Merging \eqref{eq:coi} and \eqref{eq:wb2wg}, one obtains:
\begin{equation}
  \label{eq:coi_definition}
  \wcoi(t) = \bm{\hcoi}^\top \bm{\omega}(t) + \alpha \, ,
\end{equation}
where 
\begin{equation*}
   \begin{aligned}
   \bm{\hcoi}^\top &= \bm{m}^\top \bm B_{\rm bg}^{+} [\bm B_{\rm bus} + \bm B_{\rm g}] \, , \\
   \alpha &= \bm{m}^\top \{\bm 1_m -  \bm B_{\rm bg}^{+} [\bm B_{\rm bus} + \bm B_{\rm g}]\} \, \bm 1_n \, .
   \end{aligned}
\end{equation*}
Vector $\bm{\hcoi}$ is an $n \times 1$ vector of weighting factors that depends on machine inertia constants and network topology through the imaginary part of the network admittance matrix; and $\alpha$ is a scalar offset term ensuring \(\wcoi = 1\) at nominal steady-state operating conditions.  Typically, $\alpha \ll 1$.  The reader can find more details on \eqref{eq:coi_definition} in \cite{milano_rotor_2018}.
    
Equation \eqref{eq:coi_definition} assumes that the frequency of the CoI is computed solely based on the contributions from synchronous machines, neglecting any influence from inverter-based resources (IBRs).  However, the contribution of grid-following converter is marginal (see, e.g., \cite{10253005}) and the virtual inertia of grid-forming converters can be embedded in \eqref{eq:coi_definition} in the same way as the physical inertia of conventional machines.  Thus, without substantial loss of generality,  we merge \eqref{eq:coi_bus} and \eqref{eq:coi_definition} and obtain:
\begin{equation}
\label{eq:final_complex_formula}
\boxed{\wcoi(t) = \bm{\hcoi}^{\top} \big[ \bm{H}(t) \,\dot{\bm{p}}(t) + \bm{K}(t) \, \dot{\bm{q}}(t) \big]  + \alpha}
\end{equation}

\subsection{Special Case: Simplifications of the General Formulation}

The expressions derived above can be simplified using usual assumptions for high voltage transmission systems as recommended in the IEC 60909 standard for short-circuit calculations \cite{iec60909_2016}.  Thus, the matrices approximate to:
\begin{align*}
\bm{A} &\approx \Gbus ,  &
\bm{B} &\approx -\Bbus , \\ 
\bm{C} &= \bm{B} , &
\bm{D} &= -\bm{A} .
\end{align*}

By replacing the matrix coefficients with expressions based on network parameters, $\Ybus$, we obtain:
\begin{align*}
\bm{E} &= -\Gbus \Bbus^{-1}, \\
\bm{F} &= -\Bbus + \Gbus \Bbus^{-1} \Gbus.
\end{align*}

Accordingly, the frequency sensitivity matrices become:
\begin{align*}
\bm{H} &= \left[-\Bbus + \Gbus \Bbus^{-1} \Gbus\right]^{-1}, \\
\bm{K} &= \bm{H}^{-1} \left(\Gbus \Bbus^{-1}\right).
\end{align*}

Several observations follow from these simplifications:

\begin{itemize}
\item In distribution networks, the matrices $\Bbus$ and $\Gbus$ are typically of comparable magnitude. Therefore, the two terms in the expression for $\bm{H}$ have similar contributions and have therefore to be retained.
\item Matrix $\Gbus$ is not inverted in the expressions for either $\bm{H}$ or $\bm{K}$.  This ensures numerical robustness in cases where $\Gbus \approx \bm{0}_{n,n}$, e.g., in high-voltage transmission systems.
\item If $\Gbus$ is negligible compared to $\Bbus$, $\bm{K} \approx \bm{0}_{n,n}$ and $\bm{H} \approx -\Bbus^{-1}$. In this case, equation \eqref{eq:final_complex_formula} reduces to:
\end{itemize}

\begin{equation}\label{eq:final_coi}
\boxed{\wcoi(t) = - \bm{\hcoi}^\top \Bbus^{-1}  \, \dot{\bm{p}}(t) + \alpha}
\end{equation}

This approximated expression provides a simplified, yet often sufficiently accurate, estimate of the frequency of the CoI for transmission-level systems where reactive effects can be neglected.

\subsection{Stochastic Formulation of Frequency Deviations}

To extend the previous deterministic model to a stochastic framework, we first formulate the noise of the power injections at buses as stochastic processes.  For the active power, the infinitesimal increment of the regulated active power, for example, for an Ornstein-Uhlenbeck process, is defined as:
\begin{equation}
\label{eq:pnoise}
d\bm{p}(t) = \bm a_p(\bm{p}, t) \, dt + \bm \Xi_p(\bm{p}, t) \, d\Wiener{t},
\end{equation}
with:
\begin{itemize}
    \item $\bm a_p(\bm{p}, t)$: vector of drift terms;
    \item $\bm \Xi_p(\bm{p}, t)$: matrix of diffusion terms; and
    \item $\Wiener{t}$: vector of independent Wiener processes.
\end{itemize}

Similarly, the expression of the stochastic processes of the reactive power injections can be written as:
\begin{equation}
\label{eq:qnoise}
d\bm{q}(t) = \bm a_q(\bm{q}, t) \, dt + \bm \Xi_q(\bm{q}, t) \, d\Wiener{t}.
\end{equation}

Focusing for simplicity only on the active power and integrating \eqref{eq:pnoise} yields the frequency as a stochastic process (Itô interpretation):
\begin{equation}
\label{eq:itos_deriv}
\begin{aligned}
    \bm p(t) &= \bm p(0) + \int_0^t \bm a_p(\bm{p}, s) ds 
    + \int_0^t \bm \Xi_p(\bm{p}, s) d\Wiener{s} .
\end{aligned}
\end{equation}

We can thus rewrite the expression \eqref{eq:final_complex_formula} as:
\begin{equation}
\label{eq:final_complex_formula_noise}
\boxed{d \wcoi(t) = \bm{\hcoi}^{\top} \big[ \bm{H}(t) \, d\bm{p}(t) + \bm{K}(t) \, d\bm{q}(t) \big] }
\end{equation}
and its simplified version \eqref{eq:final_coi} as:
\begin{equation}
  \label{eq:stochastic_freq}
  \boxed{d\wcoi(t) = -\bm{\hcoi}^\top \Bbus^{-1} \, d\bm{p}(t)}
\end{equation}
The two expressions above indicate that the stochastic formulation preserve the linear relationship between active power and frequency deviations of the deterministic counterparts.  Notably, they also indicate that frequency stochastic variations are function of the values of $d \bm{p}$ and $d\bm q$, not of the variations of the rate of change of these powers.  

In summary, the distribution and statistical properties of $\wcoi$ are a linear combination of the distribution and statistical properties of the noise of the active and reactive power injections, not of their time derivatives. 
The implications of this formulation, particularly regarding the statistical distribution of frequency under stochastic disturbances, are illustrated through simulations in Section \ref{sec:studycases}.

\section{Applicability of CLT in Power systems}
\label{sec:CLTdescription}

The CLT states that for independent and identically distributed (i.i.d.) random variables \(\xi_1, \xi_2, \dots, \xi_N\), each with finite expectation \(\mu\) and variance \(\sigma^2\), the normalized sum converges in distribution to a normal distribution as \(N\rightarrow\infty\):
\begin{equation}
    \frac{1}{\sqrt{N}}\sum_{i=1}^{N}(\xi_i - \mu) \xrightarrow{d} \mathcal{N}(0, \sigma^2).
\end{equation}

A generalized form, the Lindeberg Central Limit Theorem, relaxes the requirement of identical distributions. It states that for independent random variables \(\xi_{N,i}\), each with zero expectation \({\rm E}[\xi_{N,i}] = 0\) and finite variance \(\text{Var}(\xi_{N,i}) = \sigma_{N,i}^2\), the sum
\begin{equation}
    \varsigma_N = \sum_{i=1}^{N}\xi_{N,i} 
\end{equation}
converges to a normal distribution if the following Lindeberg condition is satisfied:
\begin{equation}
    \lim_{N\rightarrow\infty}\frac{1}{\varsigma_N^2}\sum_{i=1}^{N} {\rm E}\left[\xi_{N,i}^2 \mathbb{I}_{\{|\xi_{N,i}|>\epsilon \varsigma_N\}}\right] = 0,
\end{equation}
where \(\varsigma_N^2=\sum_{i=1}^{N}\sigma_{N,i}^2\), \(\epsilon>0\), and \(\mathbb{I}\) denotes the indicator function. 

In a pragmatic sense, Lindeberg condition ensures that no single random variable disproportionately influences the variance of the total sum, verifying a balanced influence among all terms to allow convergence to a normal distribution.

In power systems, the presence of multiple stochastic sources, such as loads and distributed generation, suggests potential applicability of the CLT to system-wide frequency deviations. However, as indicated in \eqref{eq:final_complex_formula_noise}, the frequency variation of the CoI explicitly depends on the stochastic variations at each bus, which are weighted by system-specific coefficients tied directly to the network's topology and impedances. 

The main reasons limiting the application of the CLT, particularly in transmission systems,  are two, as follows.
\begin{itemize}
\item \textit{Limited Number of Terms}, i.e., $N$ is small.  The CLT requires a large (theoretically infinite) number of random variables.  Although numerous stochastic sources exist at the distribution level, their influence aggregates to a limited number of interconnection buses in the transmission system, typically small compared to the requirements of the theorem. Then, the finite aggregation on the buses does not satisfy the core requirement (\(n\rightarrow\infty\)) for the CLT convergence.
\item \textit{Non-Identical Weights}, i.e., impact of network topology and different capacity of generators.  Even with a substantial number of interconnected buses, each bus contribution is scaled uniquely by coefficients given in \eqref{eq:final_complex_formula_noise}.  These ``weighting factors'' vary significantly based on network topology, impedances, and operating conditions, disrupting the homogeneity and diminishing influence required by the Lindeberg CLT.
\end{itemize}

Thus, due to the limited number of buses (finite terms) and the presence of distinct weighting factors from network parameters, the CLT cannot be directly applied to frequency variations in power systems, and the assumption of normal distribution based on CLT must be carefully reconsidered.

\begin{figure}[tbph]
    \centering
    \includegraphics[width=1\columnwidth]{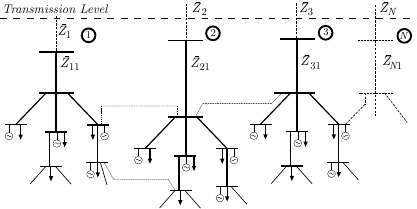}
    \caption{Stochasticity propagation -- Distribution system.}
    \label{fig:stoch_transmission1}
    \vspace{-2mm}
\end{figure}

\begin{figure}[tbph]
    \centering
    \includegraphics[width=1\columnwidth]{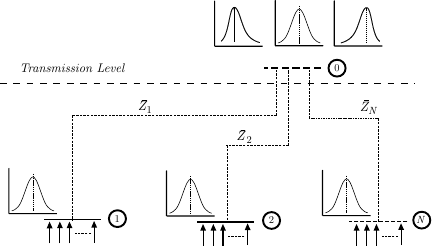}
    \caption{Stochasticity propagation -- Transmission system.}
    \label{fig:stoch_transmission2}
    \vspace{-2mm}
\end{figure}

To illustrate these constraints, consider the transmission and distribution system configurations depicted in Figs.\ref{fig:stoch_transmission1} and \ref{fig:stoch_transmission2}.  In Fig.~\ref{fig:stoch_transmission1}, buses \(1, 2, \dots, N\) represent interconnection points between distribution networks and the transmission grid. Downstream of each interconnection bus, there typically exists a complex distribution network composed of many interconnected components, often arranged in highly meshed topologies. Within such configurations, the resulting matrices $\Bbus$ and $\Gbus$ include a large number of elements, each individually small in magnitude relative to the overall network. If these elements are sufficiently similar and uniformly distributed across the network, no single branch or component significantly dominates the aggregate influence. This granular and balanced distribution of contributions should be mathematically tested through the Lindeberg theorem, thus enabling the valid application of the CLT. Consequently, stochastic fluctuations from numerous sources within distribution networks aggregate at an interconnection bus, producing an approximately normal probability distribution.

On the other hand, as illustrated in Fig. \ref{fig:stoch_transmission2}, even if each individual bus (Bus 1, Bus 2, \dots, Bus $N$) exhibits Gaussian distributions for active power injections, aggregating these distributions at a higher-level bus (Bus 0) can disrupt this normality.  This deviation occurs because each bus contribution is individually scaled by the influence of distinct impedances ($\bar{Z}_1, \bar{Z}_2, \dots, \bar{Z}_N$) and associated weighting factors.  These weights cause the combined probability distribution of the aggregated active power at Bus 0, and therefore also the frequency deviations calculated using (\ref{eq:final_complex_formula}), to move away from normality.  As a result, the aggregated distribution typically becomes skewed or develops heavy tails, as the ones typically observed in the real operation of power systems, and confirmed by the simulation results and detailed analyses presented in Section \ref{sec:studycases}.

In summary, to aggregate the power contributions of the loads is not the same as to sum them up. A simple sum, in fact, loses the information on the ``weights'' corresponding to the topology of the grid.

\section{Case Study}
\label{sec:studycases}

This section evaluates the proposed methodology using two representative case studies: a modified IEEE 14-bus test system and the all-island Irish transmission system. Time-domain simulations and stochastic analyses are carried out using the power system analysis software tool Dome \cite{domePaper}.

\subsection{IEEE 14-Bus Test System}

The IEEE 14-bus system consists of 14 buses, 20 transmission lines, and 5 synchronous generators. The total system load is approximately 2.59 pu.  We utilize this simple system to show: (i) the different level of accuracy of expressions \eqref{eq:final_complex_formula} and \eqref{eq:final_coi}; and the effect of network topology on the distribution of the frequency of the CoI.

\paragraph{Power Injections and Frequency Response}

\begin{figure}[tbph]
    \centering
    \includegraphics[width=1\columnwidth]{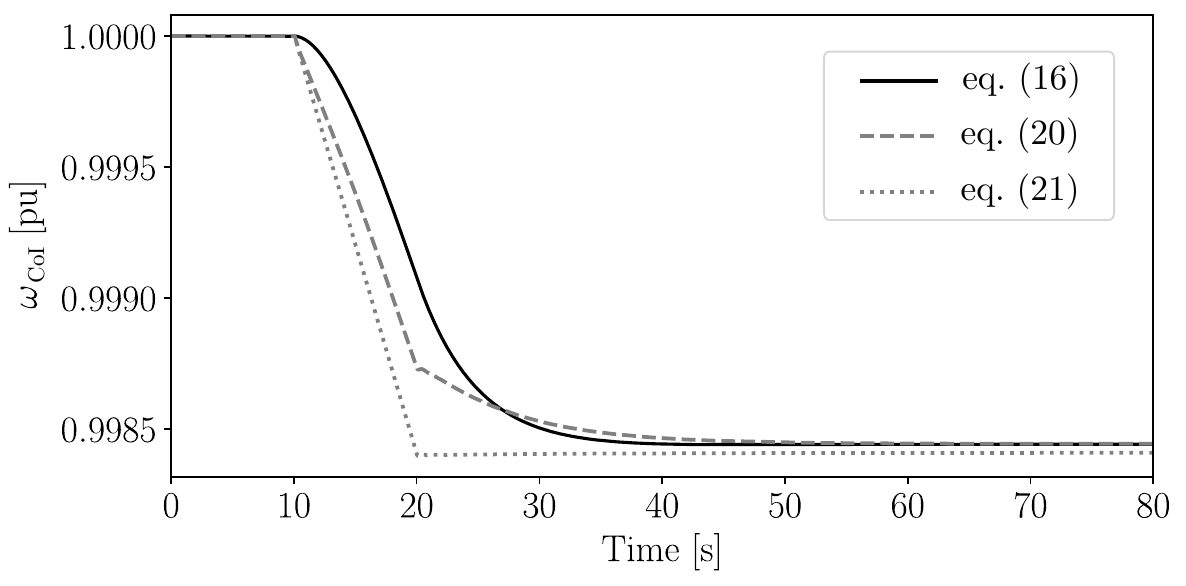}
    \caption{Calculated vs. simulated frequency response using different formulations.}
    \label{fig:omega_calculated}
\end{figure}

Figure \ref{fig:omega_calculated} shows the response of the frequency of the CoI computed using \eqref{eq:coi}, \eqref{eq:final_complex_formula} and \eqref{eq:final_coi}, following the ramp connection of a constant admittance load at Bus 4.  The load ramps up starting at $t = 10$ s, with a rate of 0.1 pu/s, over a duration of 10 s.

The frequency's expression from (\ref{eq:final_coi}) fails to properly capture transient dynamics, particularly the smooth frequency transition during load variation. 
On the other hand, the frequency computed using the full complex frequency formulation, \eqref{eq:final_complex_formula}, which incorporates both real and imaginary components, shows a significantly closer match with the simulated response.  This improved accuracy results from the model’s ability to account for both active and reactive power dynamics in a unified analytical framework.  This result suggests that voltage control, which impacts on the reactive power, not only active power control, impacts on frequency deviations and, ultimately, on system frequency quality.

\paragraph{Influence of the Grid on Stochastic Propagation}

\begin{figure}[tbph]
    \centering
    \includegraphics[width=1\columnwidth]{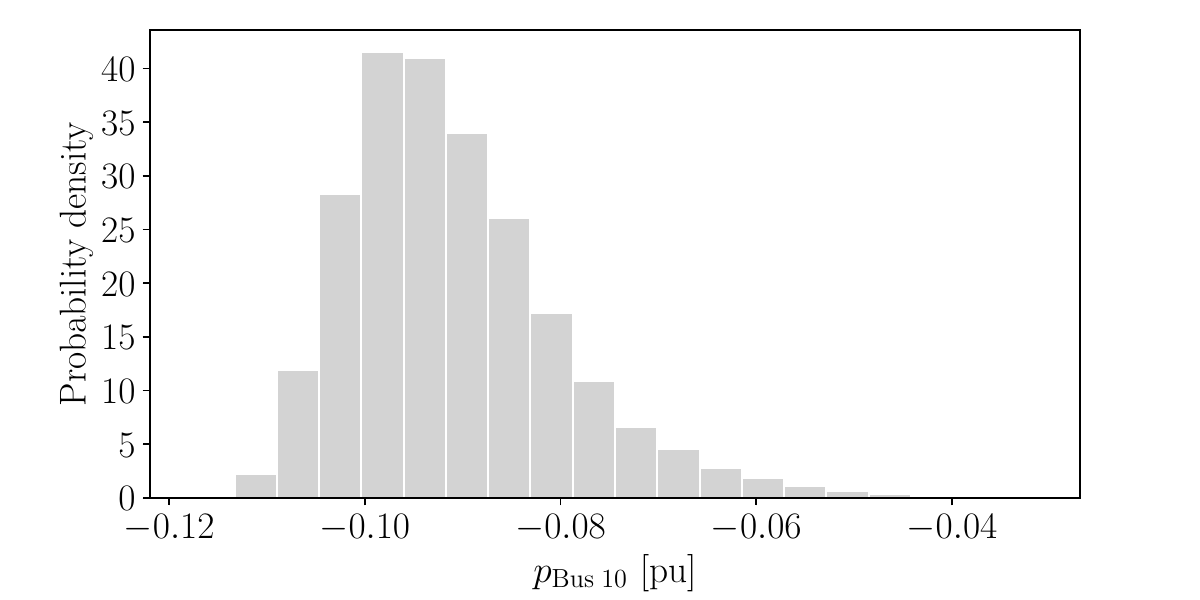}
    \caption{Weibull-distributed power injection at Bus 10.}
    \label{fig:ieee_load1_1}
\end{figure}

\begin{figure}[tbph]
    \centering
    \includegraphics[width=1\columnwidth]{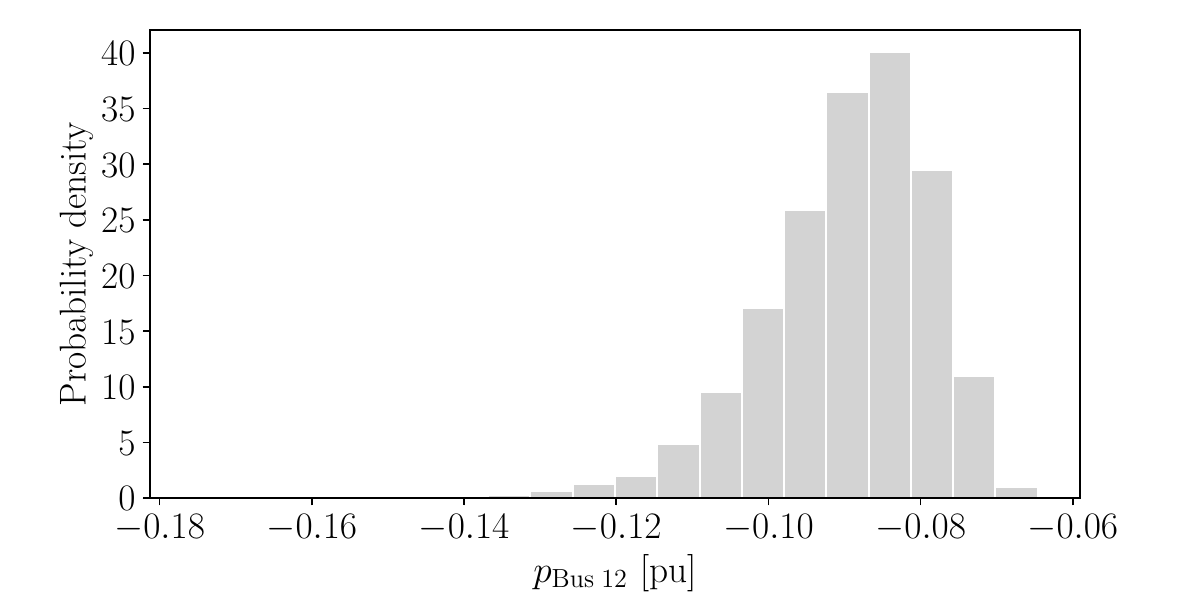}
    \caption{Weibull-distributed power injection at Bus 12.}
    \label{fig:ieee_load2_1}
\end{figure}

This second example illustrates the role of network topology.  Two stochastic loads are connected to the system: one at Bus 10 and another at Bus 12.  The noise of both loads follow a Weibull distributions but Bus 12's distribution is skewed to the left and Bus 10's to the right.  Figures \ref{fig:ieee_load1_1} and \ref{fig:ieee_load2_1} show the respective active power distributions.    The two distributions are mirrored, so that their direct sum is symmetrical.  Similar distributions are utilized for the reactive power fluctuations of these two loads.

In order to better illustrate how \eqref{eq:final_complex_formula} captures the influence of network topology, we have modified the impedances of the lines of the IEEE 14-bus system so that the two selected buses have substantially different Short-Circuit Levels (SCLs).  The SCL is defined as:
\begin{equation}
    \mathrm{SCL}_i = \frac{1}{| \bar{Z}_{ii}|},
\end{equation}
where $\bar{Z}_{ii}$ is the $i$-th diagonal element of the bus impedance matrix $\bar{\bm Z} = [\Ybus + \bar{\bm Y}_{\rm g}]^{-1}$, where $\bar{\bm Y}_{\rm g}$ is a n × n diagonal matrix that accounts for the internal admittances of the synchronous machines and step-
up transformers at generator buses.  Specifically, $\mathrm{SCL}_{10} = 4.43$ pu and $\mathrm{SCL}_{12} = 2.16$ pu.  A higher SCL indicates that the bus is strongly coupled to the rest of the network, exhibiting lower equivalent impedance and greater ability to transmit power fluctuations.  Based on this metric, thus, Bus 10 has a stronger electrical connection to the grid than Bus 12. 

\begin{figure}[tbph]
    \centering
    \includegraphics[width=1\columnwidth]{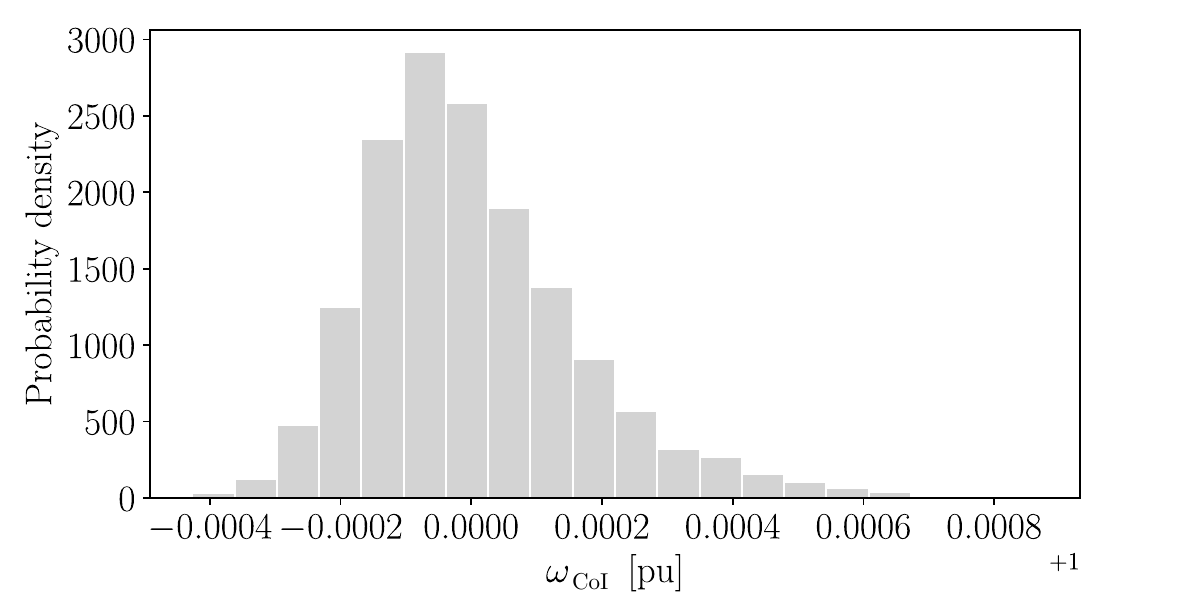}
    \caption{Resulting system frequency response influenced predominantly by Bus 10.}
    \label{fig:ieee_freq3}
\end{figure}

Figure \ref{fig:ieee_freq3} shows that the system's frequency probability density (PD) inherits the positive skew from the power injection at Bus 10, which confirms its impact on the grid has a stronger influence than the power injection at Bus 12.  The PD shown in Fig. \ref{fig:ieee_freq3} is obtained using  \eqref{eq:final_complex_formula}.  The PD obtained using the $\wcoi$ as calculated from its definition \eqref{eq:coi} are substantially the same.  

\paragraph{Applicability of the CLT in Distribution Systems}

\begin{figure}[tbph]
    \centering
    \includegraphics[width=1\columnwidth]{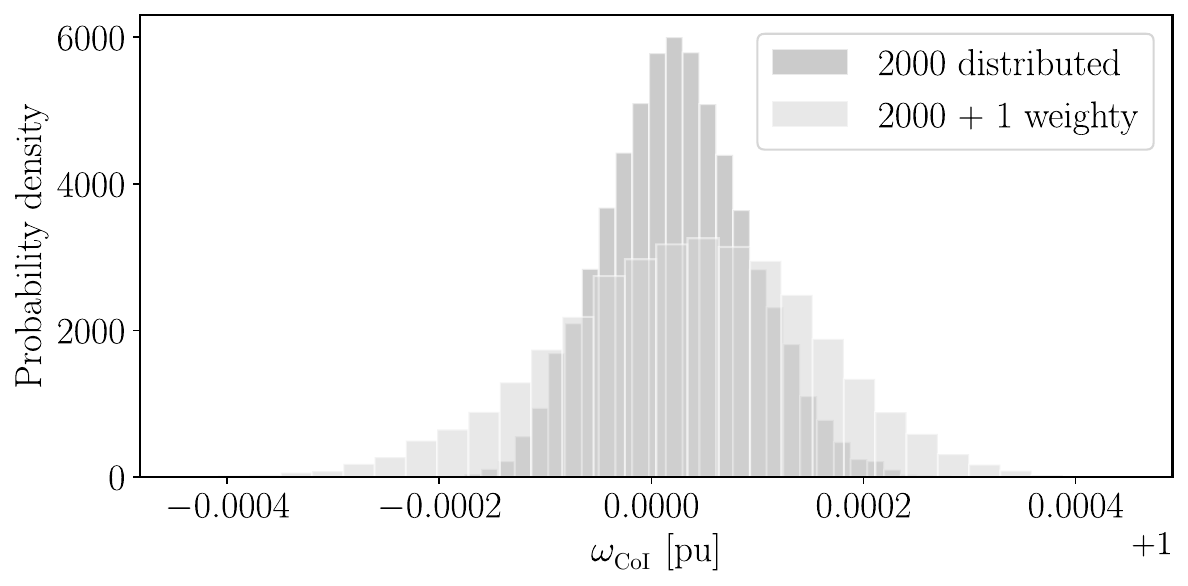}
    \caption{Frequency response on presence of uneven contributions.}
    \label{fig:distrib_uneven}
\end{figure}

To further explore the conditions under which the CLT holds in power systems, we simulate a subnetwork composed of 1,000 buses connected downstream of Bus 4. Within this subnetwork, 2,000 small loads are randomly distributed.  As shown in Fig.~\ref{fig:distrib_uneven}, the resulting frequency probability density function is close to converge toward a Gaussian shape, illustrating the effect of aggregation when contributions are numerous and relatively uniform.

However, when a single load, that represents a feeder with a big and random load attached, with disproportionately large impact is introduced into the same network, the conditions required by the Lindeberg criterion are no longer satisfied. This imbalance disrupts the convergence process and results in a frequency distribution with visible heavy tails. Figure \ref{fig:distrib_uneven} illustrates how even a single dominant term can prevent the CLT from applying, emphasizing the importance of uniformity in the magnitude of stochastic sources. As mentioned, this uniformity can be achieved indirectly when the grid is large, with small and numerous distributed loads.

\subsection{All-Island Irish Transmission System}

To evaluate the proposed methodology under realistic conditions, we apply it to a detailed model of the all-island Irish transmission system.  This model, based on publicly available data from EirGrid and SONI, includes 1,479 buses, 1,851 transmission lines, five conventional synchronous power plants, and 302 wind power plants.  Note that we have on purpose made ``weak'' the model of the system by reducing the number of machine below 7, which EirGrid and SONI strictly enforce as the minimum number of on-line machines at any given time.  This choice allows dramatizing the effect of noise and better appreciating the impact of topology on system frequency.  

In the remainder of this section, we utilize the simplified expression \eqref{eq:stochastic_freq} as the resistances of the lines of the model of the Irish system are small and thus the differences between \eqref{eq:final_complex_formula_noise} and \eqref{eq:stochastic_freq} are negligible.

\paragraph{Applicability of the CLT in Transmission Systems}

The Central Limit Theorem is often assumed to apply in distribution networks, where large numbers of small, independent sources or loads contribute to aggregate behaviors. While this assumption may appear reasonable, Fig. \ref{fig:distrib_uneven} demonstrates how a misleading conclusion can arise from naively applying the CLT at the distribution level. In the case of the transmission level, the application of CLT is even more limited because the number of buses where the stochastic loads/generators aggregates is smaller than in the distribution system, thereby degrading the conditions for applying the CLT.

\begin{figure}[tbph]
    \centering
    \includegraphics[width=1\columnwidth]{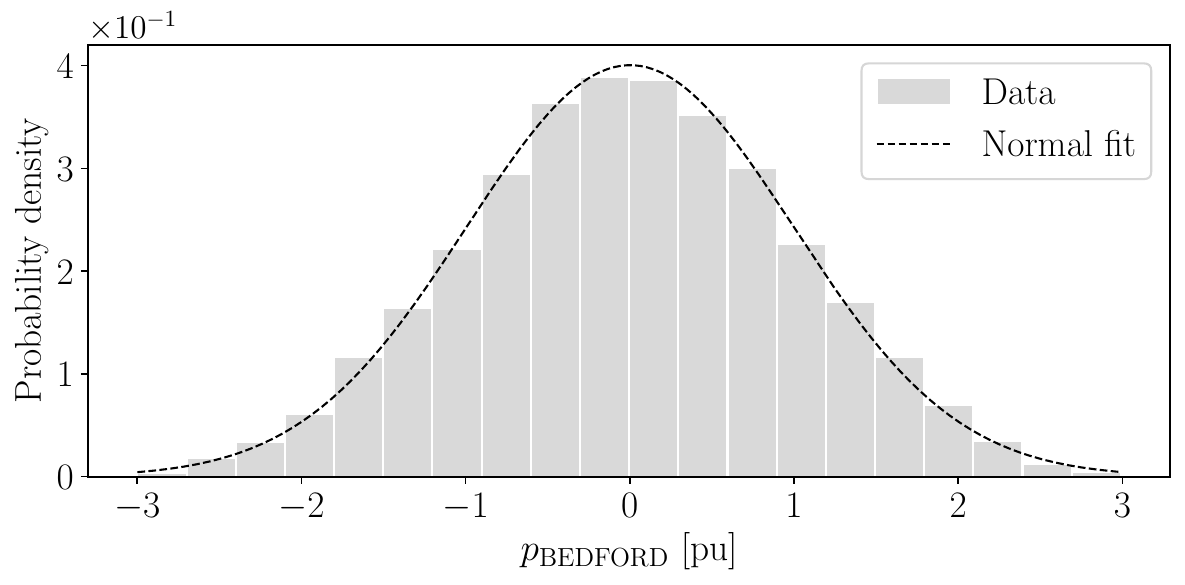}
    \caption{Aggregated active power at Bus Belford approximates a Gaussian distribution.}
    \label{fig:irish_load2}
\end{figure}

\begin{figure}[tbph]
    \centering
    \includegraphics[width=1\columnwidth]{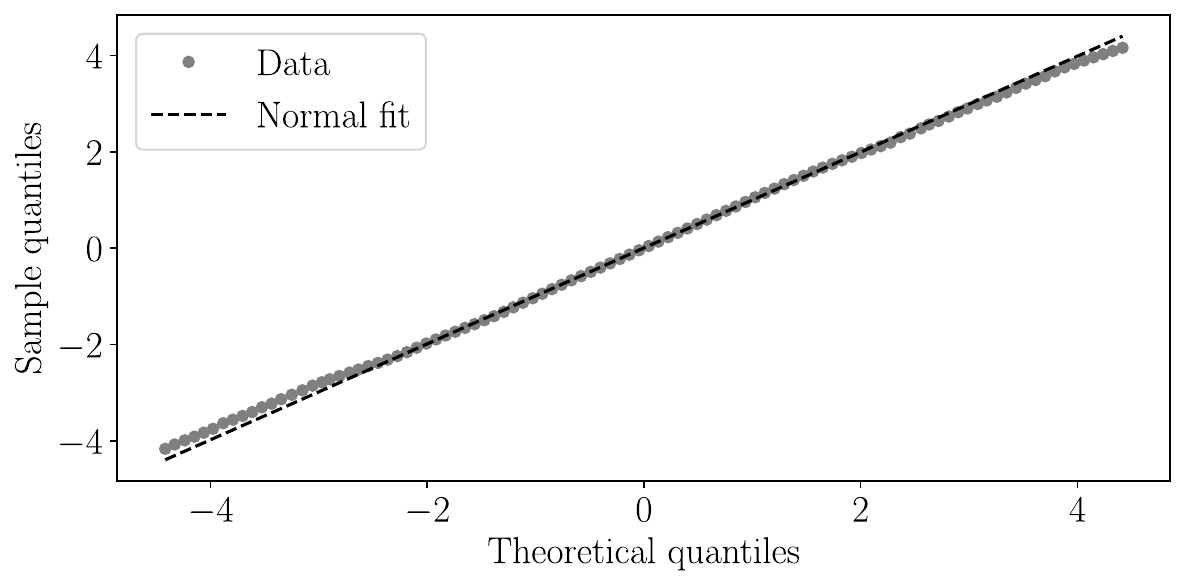}
    \caption{Normality test of the active power at Bus Belford (Q-Q plot).}
    \label{fig:irish_freq2}
\end{figure}

To illustrate this, we simulate two large groups of small stochastic loads (30,000 each) connected downstream of Bus Belford and Bus Platin. The aggregated active power at both buses approximates a normal distribution, as seen in Fig. \ref{fig:irish_load2}. This observation is supported by the quantile-quantile (Q-Q) plot shown in Fig.~\ref{fig:irish_freq2}, which confirms a close fit to Gaussian behavior for the injected power at Belford.

\begin{figure}[tbph]
    \centering
    \includegraphics[width=1\columnwidth]{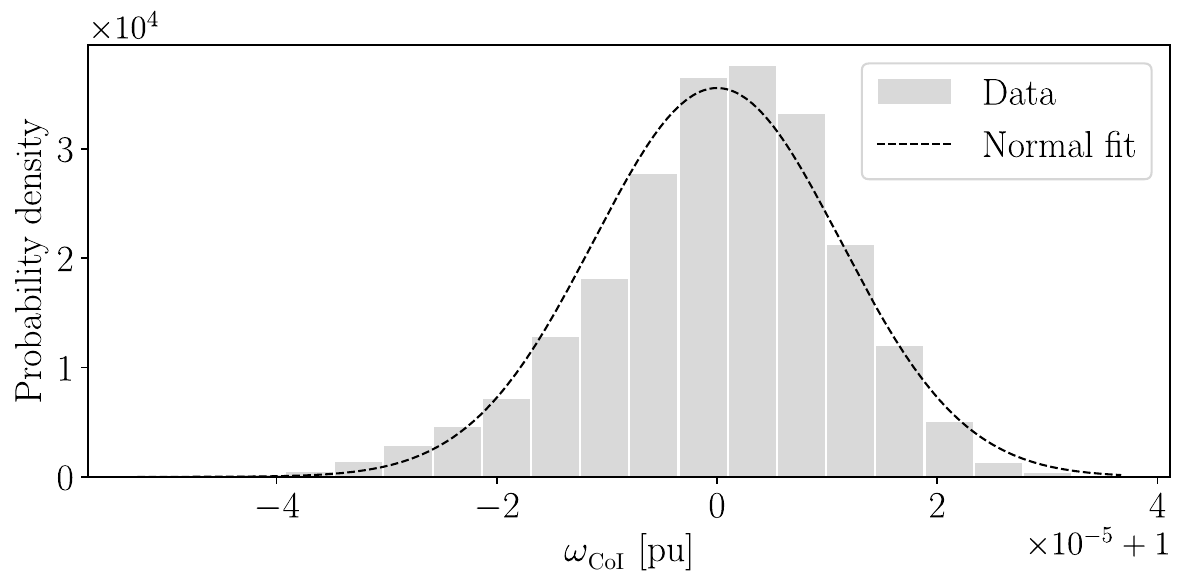}
    \caption{PD of the system frequency — non-Gaussian distribution.}
    \label{fig:irish_freq3}
\end{figure}

Despite the local application of the CLT at Bus Belford and Platin, this normality is not preserved in the system-wide frequency response. Figure \ref{fig:irish_freq3} shows the resulting PD of the system frequency, which deviates significantly from Gaussian behavior. This aligns with the analytical formulation in (\ref{eq:final_coi}), where the system frequency results from a weighted aggregation of individual power injections, shaped by network topology and bus-specific impedances.

\begin{figure}[tbph]
    \centering
    \includegraphics[width=1\columnwidth]{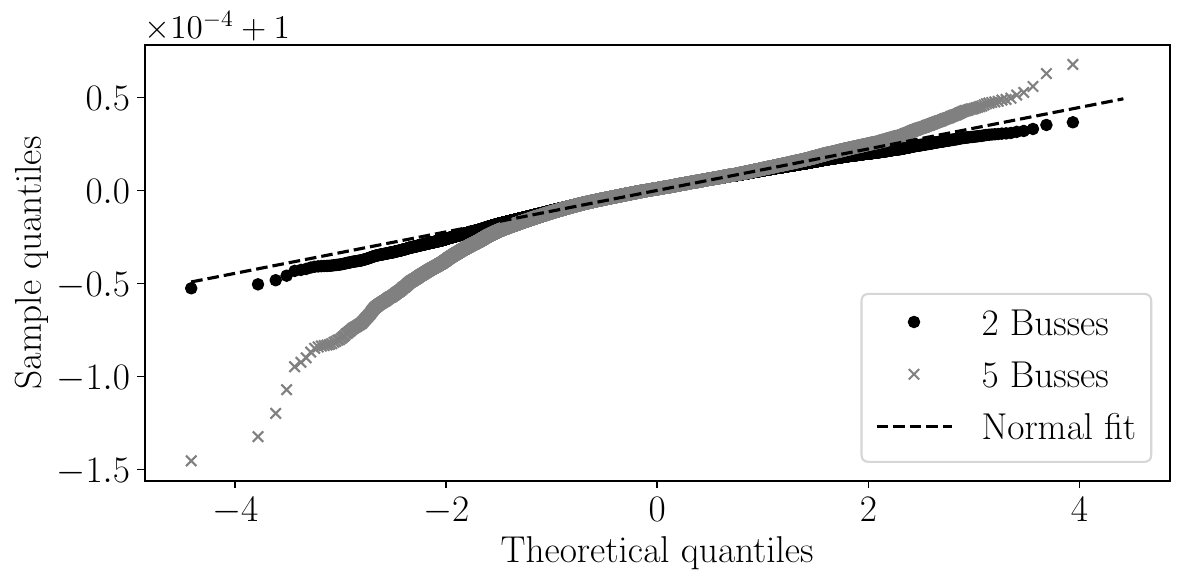}
    \caption{Normality test of the frequency distribution with 2 and 5 active buses.}
    \label{fig:irish_freq_normality}
\end{figure}

To further explore the breakdown of normality, Fig. \ref{fig:irish_freq_normality} presents a Q-Q plot to check if the frequency follows a normal distribution, under two scenarios: one with two buses (Belford and Platin), and another with five. Even with two contributing buses, the distribution shows subtle deviations from the Gaussian reference. When the number of buses increases to five, the distribution exhibits more pronounced skewness and heavier tails, which is a clear indication of non-Gaussian behavior.

These observations highlight that statistical assumptions based on the CLT may hold locally within distribution networks but break down at the transmission level due to unequal weighting, network heterogeneity, and topological effects.

It is important to mention that the stochastic perturbations used in these experiments are sufficiently small to avoid triggering frequency control mechanisms, as the frequency regulation support provided by wind power plants or automatic generation control (AGC) operations, which in the Irish system, despite the name, is operated manually.  This ensures that the frequency variations observed reflect the intrinsic transmission of stochasticity through the system, free from the influence of active regulation and the asymmetries due to the nonlinearity of wind turbine frequency control (see also the remarks in Section \ref{sub:asymmetry} below).  

In practice, large stochastic fluctuations, due for example to fault and generator tripping, such as those from aggregated wind generation or sudden load changes, are subject to correction by primary and secondary control loops.  These controllers suppress significant deviations and maintain operational frequency bounds.  However, the underlying stochastic propagation analyzed in this section remains relevant for characterizing frequency behavior under normal operating conditions to understand the impact of individual stochastic power injections, and even for designing control strategies that consider the true statistical nature of power system dynamics.

\section{Applications}
\label{sec:finalremarks}

The framework described in Sections \ref{sec:method} and \ref{sec:CLTdescription} can be utilized to tackle relevant problems that have been described in the literature.  We discuss below two of them, namely, the reconciliation of wind generation and the determination of asymmetries in power system control.

\subsection{Reconciliation of Wind Generation}
\label{sub:reconciliation}

Wind power plants are typically integrated into the grid through hierarchical, tree-like structures, as depicted in Fig.~\ref{fig:stoch_transmission1}. These topologies enable multiple levels of aggregation, from individual turbines to substation and regional levels, ultimately reaching the transmission grid. This granularity and the spatial distribution of wind farms introduce correlated stochastic behaviors that influence system-level dynamics in nontrivial ways.

Reconciliation techniques aim to resolve inconsistencies between forecast and observed wind power across aggregation levels by accounting for spatial correlation and ensuring internal consistency. Studies, such as \cite{bai_distributed_2019, sobolewski_estimation_2013}, propose frameworks to quantify and mitigate the impact of wind forecast uncertainty. The analytical framework developed in this paper complements such approaches by providing a mathematical foundation to understand how stochastic power variations originating at the distribution level propagate to the transmission system \cite{adeen_stochastic_2022}.

The proposed formulation allows the identification of conditions under which these fluctuations are either amplified or attenuated as they dissipate through the network. This becomes especially important when large-scale wind farms are connected directly to high-voltage transmission buses, where their influence on frequency dynamics is significant and should not be directly calculated by aggregation. Furthermore, due to spatial correlation among wind sources, the assumption of statistical independence,  critical for the applicability of the CLT, no longer holds. As such, standard Gaussian approximations are invalid, reinforcing the need for advanced stochastic modeling approaches that explicitly consider dependence structures, as the one presented in this paper.

\subsection{Asymmetry in Frequency Distributions}
\label{sub:asymmetry}

An increasingly observed phenomenon in modern power systems is the asymmetry of the frequency distribution, particularly in systems with high shares of inverter-based renewable generation.  As discussed in studies such as \cite{del_giudice_effects_2021} and \cite{kraljic_towards_2023}, this asymmetry can degrade system performance and reduce the effectiveness of traditional frequency control strategies.  Related bibliography attributes the effect of asymmetry to nonlinearities in system dynamics, control actions, and unbalanced contributions from heterogeneous generation sources \cite{asymmetry_2024}.  

The methodology proposed in this paper provides a mathematical rationale for the complete understanding and analysis of such asymmetries. Specifically, it reveals how frequency deviations are influenced by the different weighting of power injections, particularly from large-scale units connected at the transmission level, which cause disproportionate impacts due to their size and location. As a result, the system's aggregated stochastic response deviates from symmetric or Gaussian distributions.

This insight lays the concepts for investigating how frequency control mechanisms behave under non—Gaussian conditions, including droop control, AGC, and fast frequency response schemes. It also opens the possibility of designing improved control strategies that explicitly account for individual contributions to the asymmetries and the overall result, enhancing the resilience and frequency quality of power systems.

\section{Conclusion}
\label{sec:conclusion}

This paper introduces an analytical framework to examine the effects of stochastic power injections on frequency quality in power systems.  Using the frequency divider concept and the complex frequency formulation, we explicitly quantify how local active and reactive power fluctuations propagate through the network, affecting global frequency behavior.  A key insight from this study is that the CLT must be applied cautiously in power system analyses.  Specifically, we demonstrate that the assumption of normality in frequency deviations can fail due to the limited number of aggregated sources and the uneven contributions of individual buses.  Our analytical derivations demonstrate that network topology and impedance distributions significantly impact the resultant frequency statistics.

Simulation results validate these theoretical findings, revealing that frequency distributions may deviate from Gaussian behavior, even when local stochastic injections are normally distributed. These outcomes underline the importance of carefully considering system-specific conditions and statistical assumptions, particularly in low-inertia and renewable-rich power grids, to enhance modeling accuracy and reliability.

Future research directions include further exploration of non-Gaussian modeling approaches, detailed analysis of asymmetries in frequency distributions, and improved methods for integrating stochastic frequency modeling into power system stability and control.  We will also further investigate how reactive power control impacts on the frequency quality of low inertia systems.

\appendix

The main result of \cite{ComplexFreq} is the following expression:
\begin{align}
\label{eq:sdot}
\dot{\bar{\bm{s}}} &= \bar{\bm{s}} \circ \bar{\bm{\eta}} + \bar{\bm{S}} \bar{\bm{\eta}}^*,
\end{align}
which links the vector of complex frequency $\bar{\bm{\eta}} = \bm{\rho} + j \bm{\omega}$ to the time derivative of the vector of complex powers $\bar{\bm{s}}$ injected at network buses and to the matrix of complex power flows $\bar{\bm{S}}$ in the transmission lines and transformers.  In \eqref{eq:sdot} where $\bar{\bm{\eta}}^* = \bm{\rho} - j \bm{\omega}$ and $\circ$ denotes the element-wise product.

In this formulation, $\bar{\bm{S}} = \bm{P} + j \bm{Q}$ is a matrix of the same size of the number of network buses where the $hk$-th elements of matrices $\bm{P}$ and $\bm{Q}$ are:
\begin{align}
P_{hk} &= v_h v_k \left[ G_{hk} \cos \theta_{hk} + B_{hk} \sin \theta_{hk} \right], \\
Q_{hk} &= v_h v_k \left[ G_{hk} \sin \theta_{hk} - B_{hk} \cos \theta_{hk} \right],
\end{align}
where $G_{hk}$ and $B_{hk}$ are the real and imaginary parts of the admittance matrix $\Ybus$, $v_h$ is the voltage magnitude at bus $h$, and $\theta_{hk} = \theta_h - \theta_k$.  Then the power injections at network buses can be written as:
\begin{align}
p_h &= \sum_{k=1}^{n} P_{hk}, \qquad
q_h = \sum_{k=1}^{n} Q_{hk}.
\end{align}

The real and imaginary components of vector $\dot{\bar{\bm{s}}}$ can be written as:
\begin{align}
\dot{\bm{p}} &= \text{diag}(\bm{p}) \bm{\rho} - \text{diag}(\bm{q}) \bm{\omega} + \bm{P} \bm{\rho} + \bm{Q} \bm{\omega} \nonumber \\ 
&= \left[ \text{diag}(\bm{p}) + \bm{P} \right] \bm{\rho} + \left[ -\text{diag}(\bm{q}) + \bm{Q} \right] \bm{\omega}. \label{eq:p_rho} \\
\dot{\bm{q}} &= \text{diag}(\bm{q}) \bm{\rho} + \text{diag}(\bm{p}) \bm{\omega} + \bm{Q} \bm{\rho} - \bm{P} \bm{\omega}  \nonumber \\ 
&= \left[ \text{diag}(\bm{q}) + \bm{Q} \right] \bm{\rho} + \left[ \text{diag}(\bm{p}) - \bm{P} \right] \bm{\omega}. \label{eq:rho}
\end{align}

Solving equation \eqref{eq:rho} for $\bm{\rho}$:
\begin{equation}
\bm{\rho} = \left[ \text{diag}(\bm{q}) + \bm{Q} \right]^{-1} \left[ \dot{\bm{q}} - \left[ \text{diag}(\bm{p}) - \bm{P} \right] \bm{\omega} \right].
\end{equation}

Substituting into \eqref{eq:p_rho} yields:
\begin{align}
\dot{\bm{p}} &= \left[ \text{diag}(\bm{p}) + \bm{P} \right] \left[ \text{diag}(\bm{q}) + \bm{Q} \right]^{-1} \dot{\bm{q}} \\
&\quad - \Big( \left[ \text{diag}(\bm{p}) + \bm{P} \right] \left[ \text{diag}(\bm{q}) + \bm{Q} \right]^{-1} \left[ \text{diag}(\bm{p}) - \bm{P} \right] \\
&\quad + \left[ \text{diag}(\bm{q}) - \bm{Q} \right] \Big) \bm{\omega}.
\end{align}

Grouping terms, we can express the frequency vector as:
\begin{equation}
\boxed{\bm{\omega} = \bm{H} \, \dot{\bm{p}} + \bm{K} \, \dot{\bm{q}},}
\end{equation}
which is the derived expression (\ref{eq:coi_bus}) and where:
\begin{align}
\bm{H} &= \big \{ - \left[ \text{diag}(\bm{p}) + \bm{P} \right] \left[ \text{diag}(\bm{q}) + \bm{Q} \right]^{-1} \left[ \text{diag}(\bm{p}) - \bm{P} \right] \nonumber \\
&\quad - \left[ \text{diag}(\bm{q}) - \bm{Q} \right] \big \}^{-1}, \\
\bm{K} &= \bm{H} \left[ \text{diag}(\bm{p}) + \bm{P} \right] \left[ \text{diag}(\bm{q}) + \bm{Q} \right]^{-1}.
\end{align}

Finally, relating this formulation to the general frequency derivation of (\ref{eq:final_complex_formula}):
\begin{align}
\bm{A} &= \text{diag}(\bm{p}) + \bm{P}, \\
\bm{B} &= -\text{diag}(\bm{q}) + \bm{Q}, \\
\bm{C} &= \text{diag}(\bm{q}) + \bm{Q}, \\
\bm{D} &= \text{diag}(\bm{p}) - \bm{P} .
\end{align}


\vfill

\end{document}